\documentclass[journal=jctcce, manuscript=article]{achemso}

\usepackage{chemformula} 
\usepackage[T1]{fontenc} 
 \usepackage[colorlinks=true,linkcolor=blue,urlcolor=blue,citecolor=blue]{hyperref}
\usepackage{array}
\usepackage{lipsum}
\usepackage{bm}
\usepackage{subfigure}
\usepackage{amssymb}
\usepackage{multirow}
\usepackage{tabularx}
\usepackage{amsmath}
\usepackage{braket}
\usepackage{caption}
\usepackage{float}
\usepackage{caption}
\usepackage{enumerate}
\usepackage{ulem}

\usepackage{caption}

\renewcommand{\vec}[1]{\mathbf{#1}}

\SectionsOn
\SectionNumbersOn

\author{Zhandos Moldabekov}
\email{z.moldabekov@hzdr.de}
\affiliation{Center for Advanced Systems Understanding (CASUS), D-02826 G\"orlitz, Germany}
\alsoaffiliation{Helmholtz-Zentrum Dresden-Rossendorf (HZDR), D-01328 Dresden, Germany}

\author{Sebastian Schwalbe}
\affiliation{Center for Advanced Systems Understanding (CASUS), D-02826 G\"orlitz, Germany}
\alsoaffiliation{Helmholtz-Zentrum Dresden-Rossendorf (HZDR), D-01328 Dresden, Germany}

\author{Maximilian P. B\"ohme}
\affiliation{Center for Advanced Systems Understanding (CASUS), D-02826 G\"orlitz, Germany}
\affiliation{Helmholtz-Zentrum Dresden-Rossendorf (HZDR), D-01328 Dresden, Germany}

\author{Jan Vorberger}
\affiliation{Helmholtz-Zentrum Dresden-Rossendorf (HZDR), D-01328 Dresden, Germany}

\author{Xuecheng Shao}
\affiliation{Department of Chemistry, Rutgers University, Newark, NJ 07102, USA}

\author{Michele Pavanello}
\affiliation{Department of Chemistry, Rutgers University, Newark, NJ 07102, USA}
\affiliation{Department of Physics, Rutgers University, Newark, NJ 07102, USA}

\author{Frank~R. Graziani}
\affiliation{Institute of Theoretical Physics, University of Wroclaw, 50-204 Wroclaw, Poland}

\author{Tobias Dornheim}
\email{t.dornheim@hzdr.de}
\affiliation{Center for Advanced Systems Understanding (CASUS), D-02826 G\"orlitz, Germany}
\alsoaffiliation{Helmholtz-Zentrum Dresden-Rossendorf (HZDR), D-01328 Dresden, Germany}

\title{ Bound state breaking and the importance of thermal exchange--correlation effects in warm dense hydrogen}

\abbreviations{IR,NMR,UV}
\keywords{American Chemical Society, \LaTeX}

\begin{document}

\abstract{
Hydrogen at extreme temperatures and pressures is of key relevance for cutting-edge technological applications, with inertial confinement fusion research being a prime example. In addition,  it is ubiquitous throughout our universe and naturally occurs in a variety of astrophysical objects. In the present work, we present exact \textit{ab initio} path integral Monte Carlo (PIMC) results for the electronic density of warm dense hydrogen along a line of constant degeneracy across a broad range of densities. 
Using the well-known concept of reduced density gradients, we develop a new framework to identify the breaking of bound states due to pressure ionization in bulk hydrogen. Moreover, we use our PIMC results as a reference to rigorously assess the accuracy of a variety of exchange--correlation (XC) functionals in density functional theory calculations for different density regions. Here a key finding is the importance of thermal XC effects for the accurate description of density gradients in high-energy density systems. Our exact PIMC test set is freely available online and can be used to guide the development of new methodologies for the simulation of warm dense matter and beyond.

\begin{figure}
\center
\includegraphics[width=5cm]{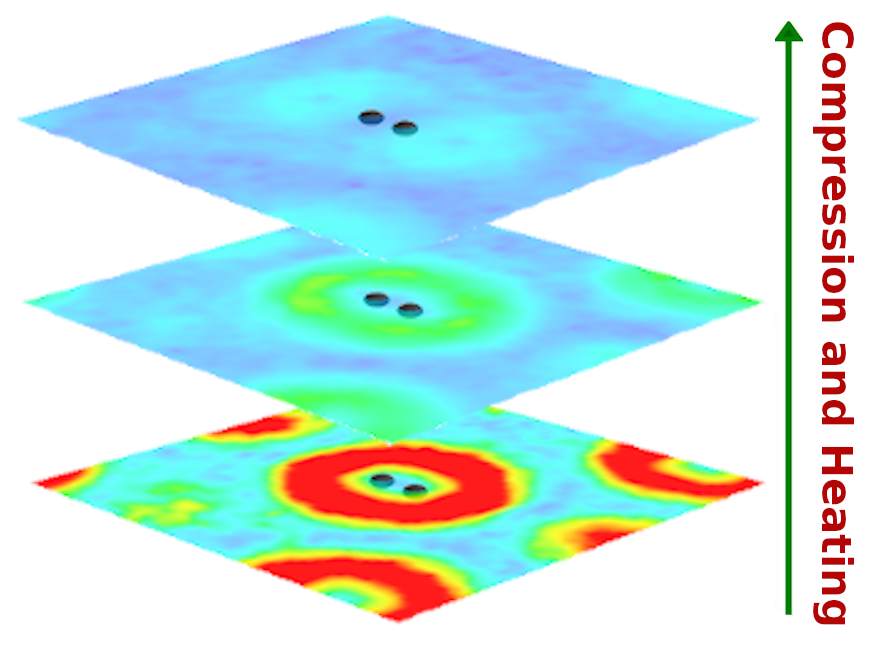}
\captionsetup{labelformat=empty}
\caption{ \label{fig:TOC} Table of Contents (TOC)/Abstract graphic. }
\end{figure} 

}

\section{Introduction}
The properties of hydrogen under extreme temperatures and pressures are of prime relevance for numerous applications in both fundamental and applied sciences.
It is ubiquitous throughout our universe and constitutes the predominant material in stars and giant planets~\cite{Benuzzi_Mounaix_2014,fortov_review}. In addition, it is of key importance for technological applications, most notably inertial confinement fusion (ICF)~\cite{hu_ICF,Betti2023}.
Despite its apparent simplicity, hydrogen offers a plethora of interesting physical effects at high-energy density (HED) conditions, including the infamous insulator-to-metal phase transition at high pressure~\cite{Pierleoni_PNAS_2010,Pierleoni_PNAS_2016} and a potential roton-type feature in the spectrum of density fluctuations in hydrogen jets~\cite{hamann2023prediction,zastrau_resolving_2014}.

A particularly important property of HED hydrogen is given by the subsequent breaking of molecular and atomic configurations with increasing densities. 
This bound state breaking plays an important role in astrophysical models, 
such as heat transport in Jupiter atmospheres through H$_{2}$ dissociation and recombination~\cite{Bell_2018_L20, Tan_2019_26}. Moreover, the breaking of hydrogen bound states at high pressures has been proposed as a possible mechanism of 
high-temperature superconductivity~\cite{Wang2022, Drozdov2015Nature}. Finally, we mention the importance of pressure ionization for the properties of a deuterium fuel capsule on its compression path in ICF experiments~\cite{hu_ICF,Fernandez2019PRL}.

As a consequence, the properties of HED hydrogen are routinely probed in experiments at large 
research facilities such as the National Ignition Facility (NIF)~\cite{Moses_NIF,Doeppner_PRL_2015}, the Omega Laser Facility~\cite{OMEGA_hydrogen}, and the Linac coherent light source (LCLS)~\cite{PhysRevResearch.5.L022023}.
Yet, the extreme conditions make the rigorous diagnostics of such experiments challenging. Therefore, 
our understanding of the physical and chemical processes at HED parameters heavily relies on simulations. In practice, the combination of a thermal Kohn-Sham density functional theory (KS-DFT)~\cite{mermin_65,wdm_book,Dornheim_review} description of the quantum degenerate electrons with a semi-classical molecular dynamics propagation of the heavier ions constitutes the most widely used simulation method to support and explain experimental works in HED science. 
It is well known that the accuracy of a KS-DFT calculation decisively depends on the utilized exchange--correlation (XC) functional, and the performance of different functionals has been analyzed extensively at ambient conditions~\cite{Cohen2008Science}.
Further, it is common practice to assess the quality of different XC functionals by benchmarking against test sets based either on 
experiments or on highly accurate theoretical results
~\cite{Goerigk_2017_32184}.

Unfortunately, the situation is substantially more difficult at HED conditions. Firstly, the construction of the first thermal XC functionals~\cite{Karasiev_T-LDA_2014,groth_prl17,dornheim_physrep18_0} that are based on finite-temperature quantum Monte Carlo (QMC) simulations~\cite{Brown_PRL_2013,dornheim_prl,dornheim_physrep18_0,Malone_PRL_2016,Schoof_PRL_2015} is a relatively recent achievement, and the development of more advanced functionals that occupy higher rungs on Jacob's ladder of functionals~\cite{doi:10.1063/1.1390175} remains in its infancy~\cite{karasiev_gga_18,Karasiev_PRB_2020,Karasiev_SCAN_2022}.
Moreover, the rigorous benchmarking of existing functionals has been hindered so far by the near complete lack of a reliable test set that would need to be based on exact simulation data for relevant properties of a real HED system.


In this work, we aim to fundamentally change this unsatisfactory situation by introducing a canonical test set for the simulation of warm dense hydrogen based on exact \textit{ab initio} path integral Monte Carlo (PIMC) calculations~\cite{Boehme_Folgepaper,Dornheim_nanoscale_2023}.
This is particularly motivated by ICF applications \cite{Zylstra2022,RevModPhys.95.025005}, where ignition---a net energy gain with respect to the energy that has been used to compress the fuel capsule---has been demonstrated in a recent milestone experiment~\cite{Betti2023}.
Here, the hydrogen capsule was shock-compressed and heated keeping electrons partially degenerate with $T\simeq T_F$~\cite{hu_ICF} ($T_F$ being the usual Fermi temperature~\cite{Ott2018}) before the final heating stage to the nuclear fusion regime takes place.
At NIF, the  compression  of the hydrogen capsule drives it from $\rho \simeq 10^{-1}~{\rm g/cc}$ to  ignition conditions with $\rho > 10^{2}~{\rm g/cc}$. During this process, the mean-inter particle distance decreases from $r_s\approx 4$  to $r_s\ll1$ (in atomic units).

Using our new \textit{ab initio} PIMC results, we study the breaking of H and H$_{2}$ bound states upon increasing the density from $r_s=4$ to $r_s=1$ along the electronic Fermi temperature $T=T_F\sim r_s^{-2}$ $(T=3.13-50.1\,$eV). This has been achieved by combining information about the electron density with  the reduced density gradient (RDG) to unambiguously identify the signature of a hydrogen bound state in bulk hydrogen.
Moreover, we provide a generalisation of the RDG that allows us to study the  interstitial electronic structure, which is of major interest for the exploration of new materials, e.g, with so-called interstitial quasiatoms \cite{Ashcroft2008PRL, Miao2020Nature}. 

Finally, we use our PIMC based test set to provide the first rigorous assessment of a variety of widely used XC functionals for KS-DFT calculations of warm dense hydrogen. In particular, we study the ability of functionals on the level of the local density approximation (LDA), generalized gradient approximation (GGA), and meta-GGA to capture the manifestation---and eventual breaking---of electron--proton bound states across a wide range of densities.
Overall, we find that the inclusion of thermal XC effects even on the level of the LDA leads to an improved description in all cases, which has important implications for the future development of improved XC functionals for warm-dense matter (WDM) applications.
Our test set is freely available online~\cite{data} and can be used to unambiguously benchmark both existing and novel tools for WDM theory.

The paper is organized as follows: In Sec. \ref{sec:QMC},  we  present new PIMC results for densities and the RDG. In Sec. \ref{s:genRDG}, we introduce a generalised RDG for the analysis of the electronic structure in different regions of the system. In Sec. \ref{s:DFTvsQMC},  we analyse the KS-DFT data computed using different XC functionals by benchmarking against the PIMC results.
In Sec. \ref{s:DFT-MD}, we demonstrate the utility of the RDG for the detection of bound state breaking in warm dense hydrogen using KS-DFT molecular dynamics (KS-DFT-MD) simulations.  We conclude the paper by emphasising our main findings and providing an outlook over potential applications of our findings.


\begin{figure*}[t!]
\center
\includegraphics[width=15cm]{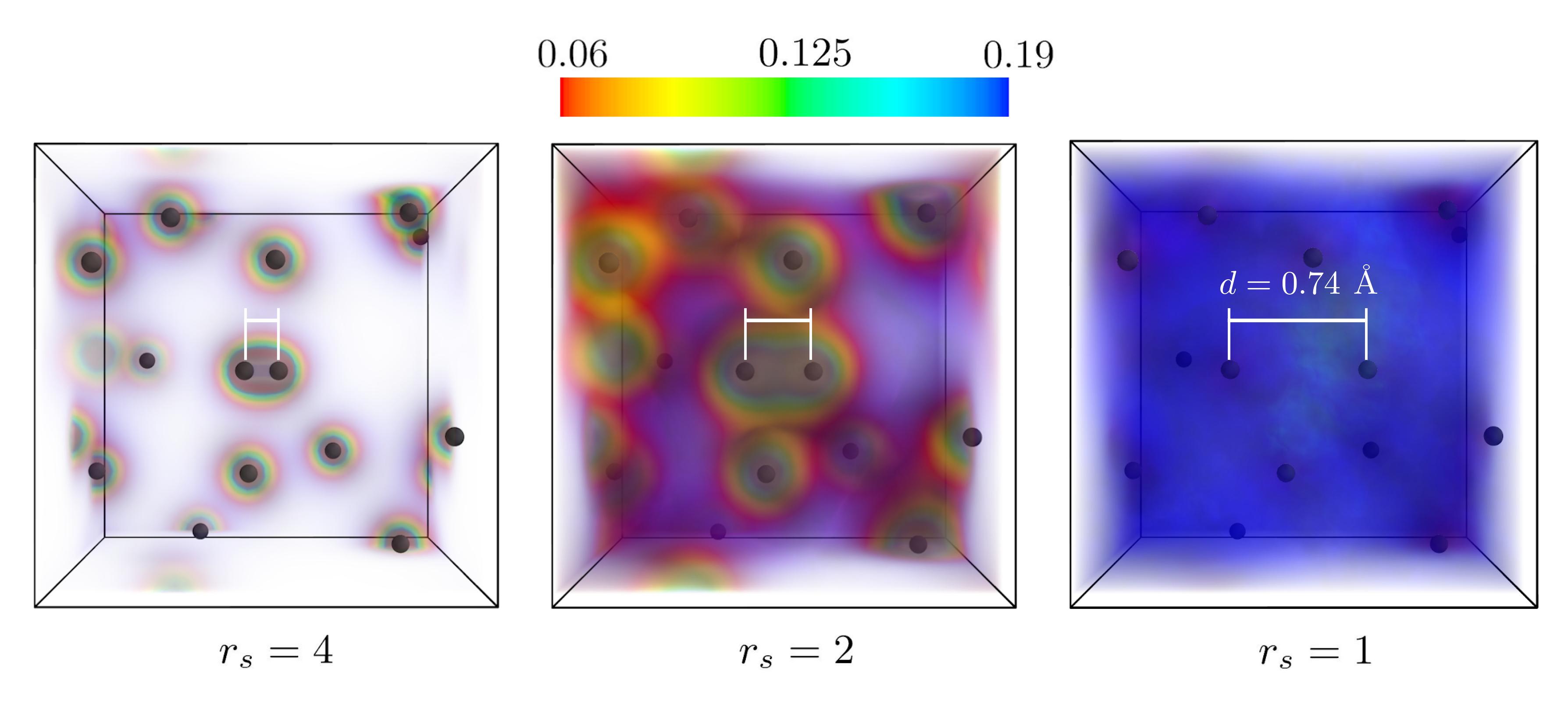}
\caption{ \label{fig:den_rs4} 
\textit{Ab initio} PIMC results for the electron density in warm dense hydrogen for a snapshot of $14$ protons at $r_s=4$ (left), $r_s=2$ (center), and $r_s=1$ (right) for $T=T_F$. In the central region of the simulation cell, two protons are positioned with a distance of $d=0.74\,$\AA$\,$  to each other (white bars); this molecular configuration is not changed for different $r_s$. 
The positions of the surrounding protons are re-scaled while keeping the angular orientation towards to the center fixed. Note that the density is normalised by the mean density $n_0$, and we cut out values above $n=0.2n_0$ for better visibility.
}
\end{figure*}

\section{Results}\label{sec:results}
\subsection{Bound state breaking in warm dense hydrogen}\label{sec:QMC}
For isolated molecular dimers, the dissociation process is theoretically studied by performing simulations at different values of the distance between the atoms constituting a molecule ~\cite{Shahi_2019_174102}; see the Supplemental Material~\cite{supplement} for additional information.
In contrast, the compression induced breaking of bound states is caused by the increasingly close average distance of the particles in bulk hydrogen. To gain insight on how this effect works for H$_{2}$ in the HED regime, we consider a disordered configuration of protons with a single H$_{2}$ molecule in the center of the simulation cell.
The corresponding PIMC results for the electronic density are shown in Fig.~\ref{fig:den_rs4}, with the left panel corresponding to the lowest considered value of the density, $r_s=4$. In addition to its relevance for ICF compression experiments, such dilute hydrogen can also be realized experimentally in hydrogen jets~\cite{zastrau_resolving_2014}. The comparably strong electron--electron coupling makes it a potentially challenging test bed for different methods, and might give rise to exotic, hitherto unobserved phenomena such as the emergence of a roton-type feature in the dynamic structure factor~\cite{hamann2023prediction}.
Due to the large interatomic distance, the H$_{2}$ molecule can easily be identified with the bare eye from the electron density at $r_s=4$.


To observe the expected change in the H$_{2}$ bond due to compression, we shrink the entire system by re-scaling the position vectors of all atoms  relative to the central H$_{2}$ molecule.
Specifically, a re-scaling of the atomic position vectors by a factor of one half and one quarter leads to the decrease of the mean-inter particle distance to $r_s=2$ and $r_s=1$, respectively. Note that the distance between the two reference atoms that make up the H$_{2}$ molecule at the center of the simulation cell always remains fixed at $d=0.74\,$\AA (white bars).

The highest considered density with $r_s=1$ is shown in the right panel of Fig.~\ref{fig:den_rs4}. In this case, $d$ is comparable to the average interatomic distance, and it is clear that the molecular bond has been broken. Indeed, all electron--proton bound states are broken in this regime as a direct consequence of the large Fermi temperature of $T_\textnormal{F}\approx50\,$eV~\cite{Hu_PRL_2017}.


A particularly interesting picture can be seen for $r_s=2$ (the center panel of Fig.~\ref{fig:den_rs4}), where we clearly observe the localisation of the electrons both around the central H$_{2}$ molecule and the surrounding hydrogen atoms and, at the same time, a stronger spreading of the electron density into the inter-atomic space compared to the dilute case of $r_s=4$. 
Such a situation is very typical for WDM~\cite{Dornheim_review},
where one cannot clearly distinguish bound and free electrons \cite{Boehme_PRL_2022}.

Comparing the electron localisation in H$_{2}$ at $r_s=2$ to that at $r_s=4$, we observe that the electronic cloud in H$_{2}$ is somewhat more extended at the higher density.
Evidently, the electronic density by itself does not provide sufficient information about the occurrence of either a molecular bond between two adjacent ions or even the formation of a proper electron--proton bound state in contrast to a screening cloud of a free electron around a nucleus.

\begin{figure*}[t!]
\center
\includegraphics[width=16cm]{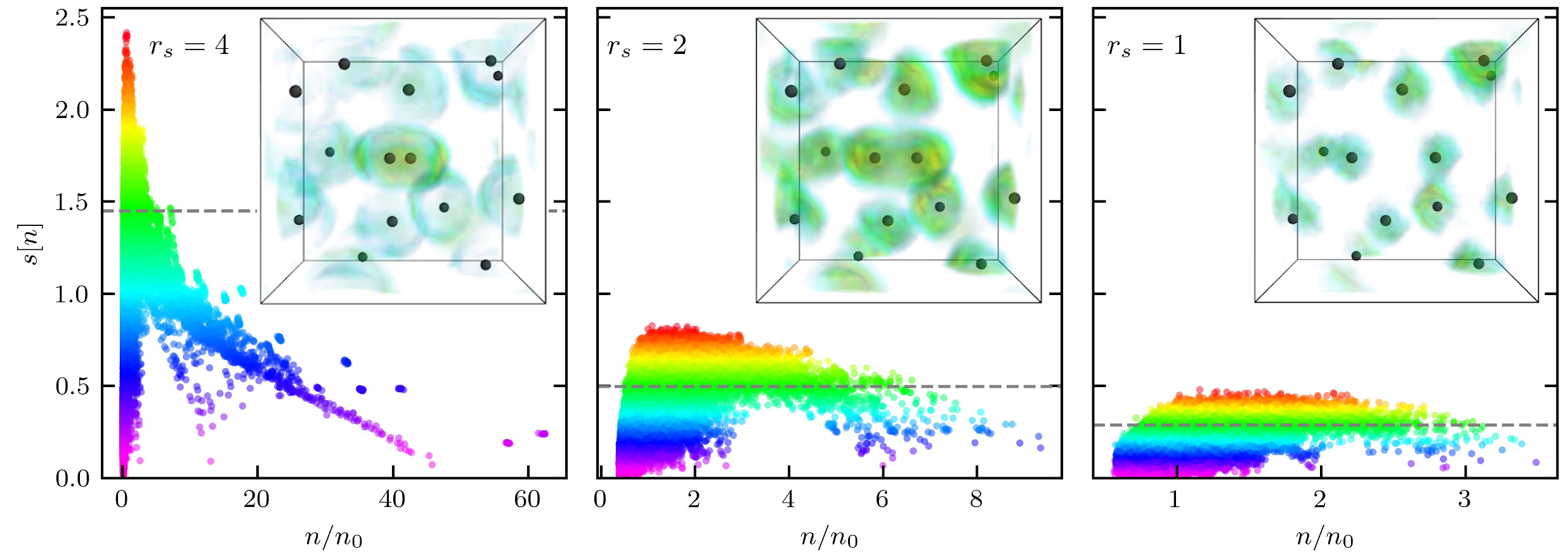}
\caption{ \label{fig:RDG_rs4} 
Distribution of the RDG [Eq.~(\ref{eq:s})] with respect to density for the warm dense hydrogen system shown in Fig.~\ref{fig:den_rs4}. 
The subplots show the corresponding distribution of the RDG in real space. 
More specifically, we show the RDG values above the horizontal dashed line in the $s[n]$ plot. This allows us to identify regions where density gradients are particularly important.
The colors above the horizontal dashed line serve as the color bar for the distribution of the RDG in real space.
}
\end{figure*} 


To overcome this diagnostic bottleneck, we use the dimensionless reduced density gradient (RDG) measure~\cite{Zupan1997JCP},
\begin{equation}\label{eq:s}
    s\left[n\right]=\frac{ \vert \vec \nabla n(\vec r) \vert}{2q_F(\vec r) n (\vec r)},
\end{equation}
where $n(\vec r)$ is the density, and $q_F(\vec r)=(3\pi^2 n(\vec r))^{1/3}$ it the local Fermi wavenumber.
The RDG has been shown to be an effective tool to identify bonding at ambient conditions~\cite{Johnson_2010_6498}. Secondly, the RDG has an appealing physical meaning as the local representation of the electronic momentum~\cite{Hugo2008JCP} within quantum kinetic energy. In particular,
$s[n]$ is the ratio of the local electron momentum to twice the local Fermi momentum  $2q_F[n]$. This ratio also naturally appears in the density gradient expansion of XC functionals in KS-DFT \cite{Perdew1986PRB, Burke1998Springer, Lucian2011PRL} and  of kinetic energy functionals in orbital-free DFT \cite{Perdew1992PLA, ke_test_jctc, zhandos_pop18, Moldabekov_cpp_2017}. In fact, $s[n]$ is a key ingredient to the construction of XC functionals beyond the local density approximation. Finally, we note that the reduced density gradient is related to the ionization potential of atoms~\cite{March1997MolecularPhysics}, which is an input quantity in multi-scale simulation models of ICF~\cite{RevModPhys.95.025005}.

In Fig.~\ref{fig:RDG_rs4}, we show the RDG $s[n]$ as a function of the electronic density  at $r_s=4$ (left), $r_s=2$ (center), and $r_s=1$ (right). In addition, we provide the corresponding distribution of $s[n]$ within the $3D$ simulation cell in the subplots. Clearly, the dependence of the RDG on the density is non-monotonic. At large density values $n\gg n_0$ (with $n_0=3/(4\pi r_s^3)$)---i.e., in close proximity to the protons---$s[n]$ decreases with the increase in the density since the Fermi momentum scales as $q_F[n]\sim n^{1/3}$ while  the  density gradient amplitude $\vert \nabla n \vert$ is limited by Kato’s cusp condition~\cite{March1997MolecularPhysics}. In the opposite limit of low densities, the density gradient $\vert \nabla n \vert$ also decreases since the Coulomb field is effectively screened in the inter-atomic space. The RDG $s[n]$ attains its largest values at a distance to a proton for which the local electron quantum momentum exceeds the Fermi momentum. To identify this region, we spatially resolve the top $40\%$ values of $s[n]$ in the subplots. Specifically, these values correspond to the data points above the horizontal dashed-line in the respective $s[n]$ plots on the left.

At $r_s=4$, we observe that a sharp peak in $s[n]$ (with $\rm max (s[n])\simeq 2.4$ ) at small $n$ identifies the electronic shell structure of a bound state both of a hydrogen atom and, in particular, in the H$_{2}$ molecule. 
Indeed, this shell structure is characteristic for an isolated H$_{2}$ molecule~\cite{supplement}, and
we have confirmed this observation by performing KS-DFT simulations with the Fermi-L{\"o}wdin orbital self-interaction correction~ \cite{Pederson_2014_121103,Pederson_2015_064112,Pederson_2015_153,Yang_2017_052505} for a single molecule at different nuclei separations; see the Supplemental Material~\cite{supplement} for additional information.
These calculations 
further substantiate our identification of the sharp peak in $s[n]$ for low $n$ with a signature of bound states in both H and H$_{2}$.

Upon increasing the density to $r_s=2$, we observe a significant flattening of the $s[n]$ distribution. Moreover, the maximum of $s[n]$ is reduced by more than a factor of two to  $\rm max (s[n])\simeq 0.8$, and the sharp signal of bound states disappears entirely. This constitutes a clear observation of the breaking of bound states in warm dense hydrogen, both for the disordered atoms and the single molecular configuration around the center. Further, it can be seen clearly that the distribution of $s[n]$ in real space does not have a clear shell structure. Finally, we remark that any qualitative difference between H and H$_{2}$ disappears at $r_s=2$, which, too, is a direct consequence of the disappearance of the electronic bound states.

In the right panel of Fig.~\ref{fig:RDG_rs4}, we show corresponding results for the highest considered density, $r_s=1$.
Here the main trend is given by a further flattening of the distribution of $s[n]$, and a further overall decrease of its magnitude with  $\rm max (s[n])\simeq 0.4$.
Looking at the depiction of $s[n]$ in real space in the subplot, we find an even more clear suppression of any electronic shell structures compared to the previous case of $r_s=2$.
Similar changes in the distribution of $s[n]$ are also observed when we keep the density fixed and increase the temperature (see the Supplemental Material~\cite{supplement}).
This indicates that $s[n]$ can be used to study bound states breaking in other ionization scenarios as well.


We note that the analysis presented in Fig.~\ref{fig:RDG_rs4} has been carried out for a set of comparably small synthetic snapshots. A corresponding RDG analysis of real bulk hydrogen based on DFT-MD simulations of $N=500$ hydrogen atoms is shown in Fig.~\ref{fig:rdg_N500} below and leads to the same trends.

\subsection{Generalised dimensionless RDG}\label{s:genRDG}

The dimensionless RDG $s[n]$ plays an important role in the construction of XC functionals beyond LDA~\cite{Zupan1997JCP, Burke1998Springer}.
To quantify the quality of thermal KS-DFT for the description of electronic density gradients, we analyse the deviation in $s[n]$ between KS-DFT and PIMC data.
In this context, we note that---as we have seen earlier---$s[n]$ attains a maximum around the outer layer of an atom or molecule, and rapidly decays both in the inter-atomic region and around the protons.
Hence, the usual definition of $s[n]$ 
puts the focus on certain density regions, which potentially limits its utility as a benchmark for the quality of different XC functionals. 
To avoid this undesirable feature, we introduce a density re-weighting of the form
$s[n]\left(n/n_0\right)^{\alpha}$, thereby generalising the dimensionless RDG defined above.
For example, by setting $\alpha=1/3$, we  compensate for the fast decay of $s[n]$ with density at $n/n_0\gg1$ due to the $q_F\sim n^{-1/3}$ dependence in Eq.~(\ref{eq:s}). 
Exploring the behavior of $s[n]\left(n/n_0\right)^{\alpha}$, we find that the variation of $\alpha$ leads to a shift in the maximum of  $s[n]\left(n/n_0\right)^{\alpha}$ into the inter-atomic region for $\alpha<0$ and closer to protons for $\alpha>0$ compared to the usual case of $\alpha=0$. In practice, the modified RDG thus allows us to perform a scanning of the quality of density gradients across different regions.
For benchmarking purposes, it is convenient to condensate the information about the generalised RDG into a single scalar number by defining the integrated measure 
\begin{equation}\label{eq:integrated}
S(\alpha)=\frac{1}{N}\int {\rm d} n ~s[n] \left(\frac{n}{n_0}\right)^{\alpha},
\end{equation}
with $N$ being the total number of particles in the integration volume. Eq.~(\ref{eq:integrated}) thus plays a central role for our assessment of a variety of XC functionals in thermal KS-DFT simulations below.

\begin{figure}[t!]
\center
\includegraphics[width=7.5cm]{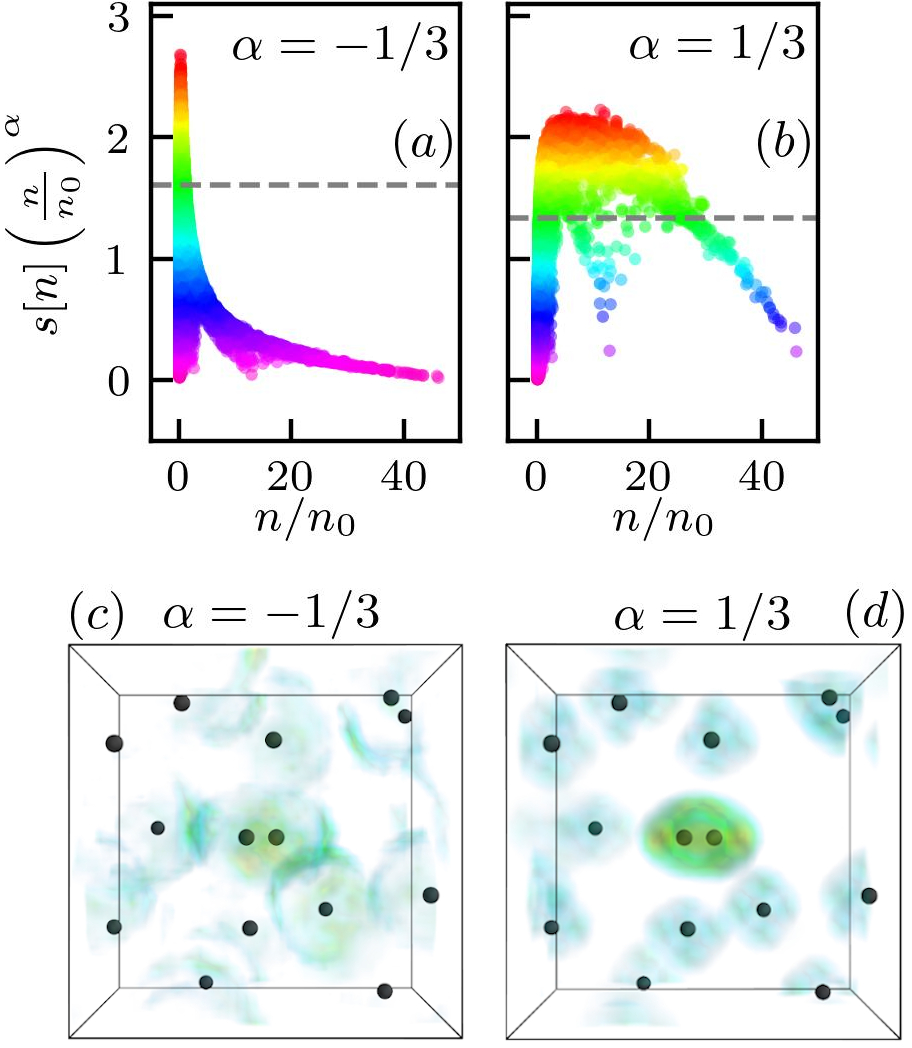}
\caption{ \label{fig:s_alpha} 
Generalized dimensionless RDG $s[n]\left(n/n_0\right)^{\alpha}$ at $\alpha=-1/3$ and $\alpha=1/3$ for  $r_s=4$ and $T=T_\textnormal{F}$. We show that the variation of $\alpha$ allows emphasizing the rich behaviour of the density gradient at different distances from protons. 
Panels (a) and (b) show the distribution of $s[n]\left(n/n_0\right)^{\alpha}$ as a function of the density at $\alpha=-1/3$ and $\alpha=1/3$, respectively. Panels (c) and (d) show corresponding results for $s[n]\left(n/n_0\right)^{\alpha}$ in real space, including data points above the dashed horizontal lines in panels (a) and (b). 
The colors above the horizontal dashed line in panels (a) and (b) serve as the color bar for the distribution of the RDG in real space in panels (c) and (d).
}
\end{figure}

In Fig.~\ref{fig:s_alpha}, we show $s[n]\left(n/n_0\right)^{\alpha}$ for $\alpha=1/3$ and $\alpha=-1/3$  computed from our exact PIMC results for hydrogen at $r_s=4$.
Comparing these data with the results for $\alpha=0$ shown in the left subplot of Fig.~\ref{fig:RDG_rs4} above, we find that setting $\alpha$ to negative values leads to a shift of the maximum of $s[n]\left(n/n_0\right)^{\alpha}$ to smaller densities, i.e., to the inter-atomic region. 
Moreover, the distribution $s[n]$ becomes substantially narrower.
In contrast, setting  $\alpha=1/3$ shifts 
the maximum to larger densities, and the RDG distribution is broadened. 
To get a more intuitive picture of the effects of the density re-weighting, we show the corresponding distribution of the generalised RDG in panels (c) and (d). Indeed, setting $\alpha=-1/3$ reveals a hitherto hidden but remarkably rich structure of density gradients between the protons, whereas $\alpha=1/3$ puts the focus on the close vicinity of the latter.


\begin{figure}
\center
\includegraphics[width=7.7cm]{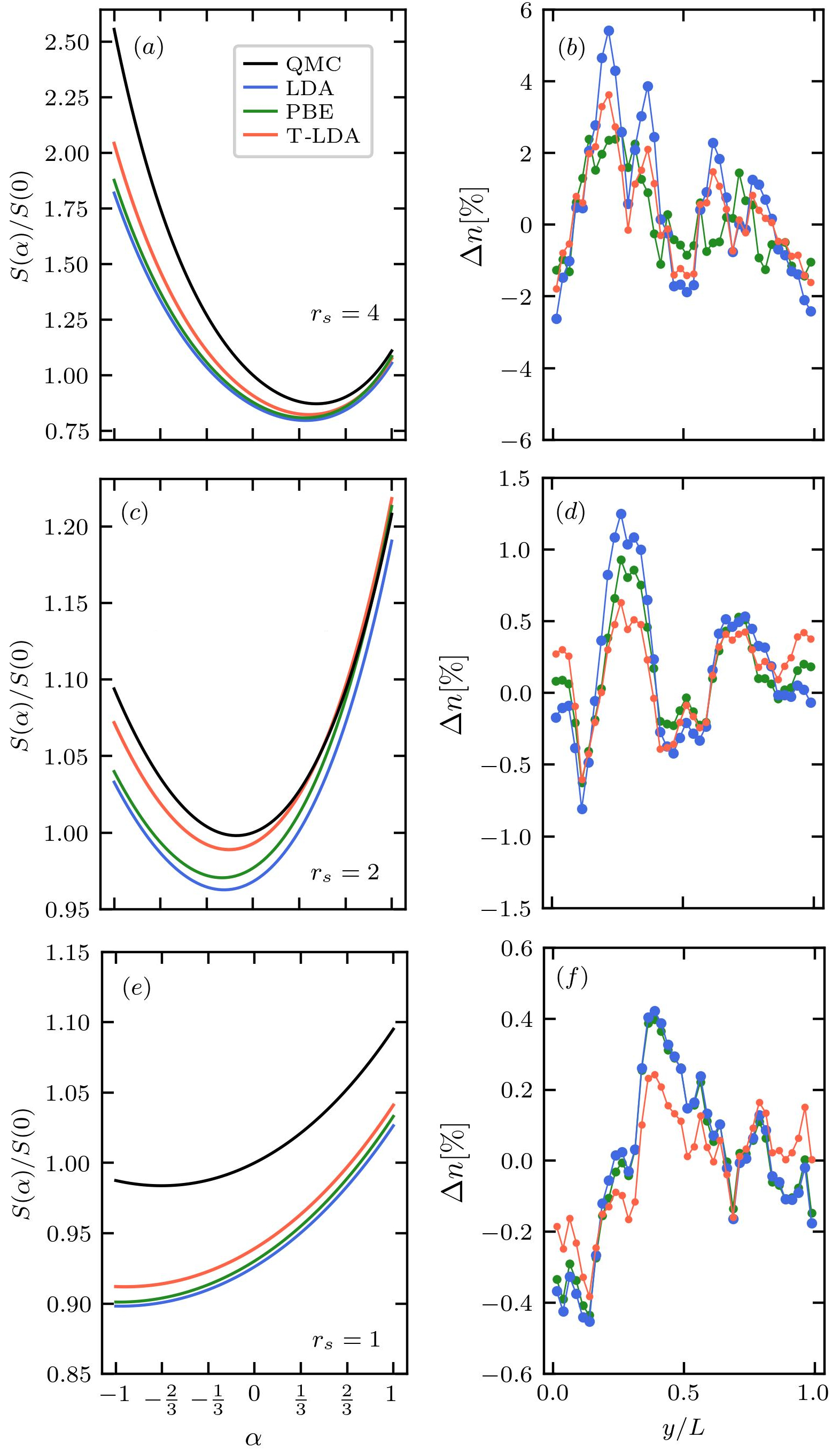}
\caption{ \label{fig:S/S0} 
 Integrated RDG measure $S(\alpha)$ [Eq.~(\ref{eq:integrated})], for (a)  $r_s=4$, (c)  $r_s=2$, and (e)  $r_s=1$. All $S(\alpha)$ curves are normalised to the result of $S(0)$ from PIMC; we find $S(0)\simeq0.913$ for $r_s=4$, $S(0)\simeq 0.344$ for $r_s=2$, and $S(0)\simeq 0.176$ for $r_s=1$.  
In addition, we show the deviation of KS-DFT results using ground-state LDA, PBE, and thermal LDA from the exact PIMC reference data for the electronic density projected along the $y$-axis in panels (b), (d), and (f).
}
\end{figure} 

\begin{figure*}
\center
\includegraphics[width=16cm]{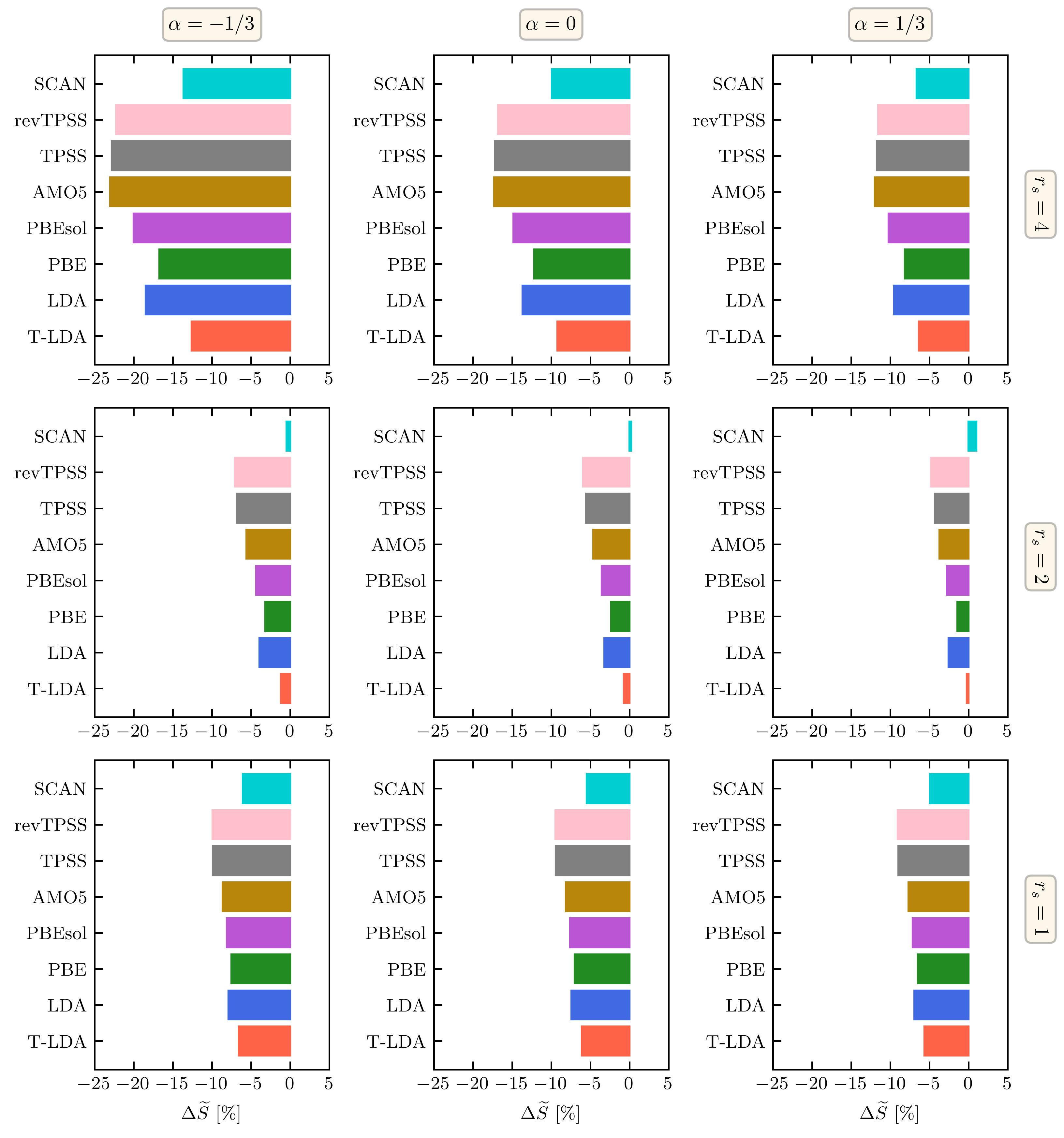}
\caption{ \label{fig:Delta_S} 
Relative difference in the integrated RDG measure [cf.~Eq.~(\ref{eq:Delta_S})] between KS-DFT simulations using different XC functionals and the exact PIMC reference data. 
The top, center and bottom rows correspond to $r_s=4$, $r_s=2$, and $r_s=1$, and the left, center, and right columns have been obtained for $\alpha=-1/3$, $\alpha=0$, and $\alpha=1/3$.
}
\end{figure*} 


In the context of the present work, we use $s[n]\left(n/n_0\right)^{\alpha}$ to test the quality of the KS-DFT results for the density gradients by comparing with the exact PIMC data. Furthermore, our results indicate that $s[n]\left(n/n_0\right)^{\alpha}$  with $\alpha<0$ is particularly useful for the analysis of the features of the electronic structure in the inter-atomic region. 

\subsection{Benchmarking XC functionals and the role of thermal effects}\label{s:DFTvsQMC}


The integrated RDG measure $S(\alpha)$, cf.~Eq.~(\ref{eq:integrated}) above, constitutes a convenient scalar quantity to analyse the capability of different XC functionals to capture density gradients, and the rich physics they entail, in real WDM systems.
We analysed the quality of eight XC functionals commonly used in material science, including the LDA as it has been parametrized by Perdew and Zunger~\cite{Perdew_LDA},  the GGA level functionals PBE~\cite{PBE}, PBEsol~\cite{PBEsol} and the Armiento–Mattsson functional (AM05)~\cite{PhysRevB.72.085108}, and the meta-GGA approximations TPSS \cite{TPSS}, revTPSS \cite{revTPSS}, and SCAN~\cite{SCAN}. 
It is important to note that these functionals have exclusively been constructed for applications at ambient conditions, where the electrons are in their respective ground state. However, using such a functional for applications in the WDM regime can potentially lead to substantial inaccuracies, as it has been reported by independent groups~\cite{kushal,karasiev_importance,karasiev_gga_18,Karasiev_SCAN_2022,Sjostrom_PRB_2014}. To rigorously assess the importance of thermal XC effects that are, by definition, not included into the aforementioned ground-state functionals, we also 
study the finite-temperature LDA (T-LDA) XC functional by Groth \textit{et al.}~\cite{groth_prl17}. 
For completeness, we note that more sophisticated thermal XC-functionals on the level of GGA and meta-GGA have been developed very recently~\cite{Karasiev_PRL_2018,Karasiev_PRB_2020,Karasiev_SCAN_2022}. Applying such novel functionals to the present canonical test set of warm dense hydrogen constitutes an important project for future work.



In the left column of Fig.~\ref{fig:S/S0}, we compare our exact PIMC reference results (black) for $S(\alpha)$ to analogous KS-DFT results based on ground-state LDA (blue), PBE (green), and thermal LDA (red); the top, center, and bottom rows correspond to $r_s=4$, $r_s=2$, and $r_s=1$.
Note that all curves are normalised to the result for $S(0)$ of PIMC at a given $r_s$.

Let us postpone the discussion of $S(\alpha)$ for now and proceed to the right column of Fig.~\ref{fig:S/S0} that shows the deviation of KS-DFT simulations to the PIMC results in the electron density projected along the $y$-axis. 
Evidently, all considered XC functionals exhibit a similar level of accuracy with deviations of a few percent at $r_s=4$ and less than one percent at $r_s=1$. 
On average, thermal LDA performs slightly better than the ground state LDA and PBE for the two highest densities, but similar to PBE at $r_s=4$.
Therefore, we cannot clearly resolve the role of thermal XC effects from this analysis.

In contrast, our results for $S(\alpha)/S(0)$ that are shown in the left column of Fig.~\ref{fig:S/S0}
reveal a substantial, systematic improvement due to the thermal LDA functional over a wide range of $\alpha$-values for all considered densities. 
This clearly demonstrates the role of thermal effects in capturing the correct RDG.

From a physical perspective, the inverted dome shape of $S(\alpha)$ is a specific feature of bulk systems. 
In the Supplementary Material~\cite{supplement}, we show that $S(\alpha)$ decays exponentially for large $\alpha$ in the case of  isolated atoms and molecules due to the decay of the electron density at large distances to the nuclei. 
For the case of warm dense hydrogen that we consider in the present work, we observe that the value of $\alpha=\alpha_{\rm min}$ at which $S(\alpha)$ attains its minimum 
changes its sign with the breaking of bound states with increasing density.
Our PIMC and KS-DFT calculations for $r_s=1$, $r_s=1.5$, $r_s=2$, $r_s=2.5$, $r_s=3$, $r_s=3.5$ and $r_s=4$ 
show that  $\alpha_{\rm min}$ changes its sign around $r_s=2.5$ when $s[n]\lesssim 1$ (see the Supplemental Material~\cite{supplement}). 
The possibility that this finding constitutes a universal sign of bound state breaking in monatomic materials is a very interesting route for future studies.


To get a broader and more complete picture of the performance of a variety of XC functionals on various rungs of Jacob's ladder of functional approximations, we consider the three representative cases of $\alpha=-1/3$, $\alpha=0$, and $\alpha=1/3$ in Fig.~\ref{fig:Delta_S}.
More specifically, we analyse the difference in $S(\alpha)$ between the exact PIMC benchmark results and the different KS-DFT data sets via 
\begin{equation}\label{eq:Delta_S}
    \Delta S(\alpha)=\frac{S_{\rm DFT}(\alpha)-S_{\rm PIMC}(\alpha)}{S_{\rm PIMC}(\alpha)}\times 100~\% \, ,
\end{equation}
which we visualise as a bar plot. 
This leads us to the following conclusions:
\begin{enumerate}[(i)]
    \item thermal LDA performs better than ground-state LDA and all other considered GGA level functionals (PBE, PBEsol, AM05);
    \item ground state PBE improves the quality of the RDG description compared to ground-state LDA;
    \item comparing the results for meta-GGA functionals with each other, we see that SCAN performs significantly better than TPSS and revTPSS;
    \item thermal LDA and SCAN provide results for the RDG of a similar quality.
\end{enumerate}
Despite the fact that SCAN is built on top of the ground-state LDA, the observed
similarity between the quality in the RDG to that of thermal LDA 
might be a result of the information about the orbital kinetic energy densities in SCAN that automatically take into account the thermal spreading of the occupation numbers.  
The latter is due to the exact constraints based design of SCAN~\cite{SCAN} utilising  the orbital kinetic energy densities $\tau =\sum_{i}^{Nb} (1/2) f_i \left|\psi_{i} \right|^2$, where $N_b$ is the number of bands, $f_i$ is the occupation number of a Kohn-Sham orbital $\psi_{i}$  at a given temperature.
Taking into account the definition of the kinetic energy operator as a double gradient in coordinate space, it is then not surprising that the two considered XC-functionals that include thermal kinetic XC-effects best describe the RDG.


In general, we observe that the quality of the RDG from KS-DFT simulations is worse at $r_s=4$ compared to $r_s=2$ and $r_s=1$. 
This is expected since the degree of electron localization around the protons is much stronger at low densities. From Fig.~\ref{fig:Delta_S}, we see that the quality of the RDG from thermal KS-DFT with the thermal LDA and ground-state SCAN functionals is extremely good at $r_s=2$. 
This is relevant information for WDM research since WDM is often generated using metals~\cite{Doeppner2023Nature} in experiments.
Interestingly, the quality of the analysed thermal KS-DFT results for the RDG is somewhat worse at $r_s=1$ compared to $r_s=2$ despite the significantly less pronounced electronic structure in this regime. 
This is a direct and somewhat artificial consequence of the small variations in the amplitude of the RDG. While being a subtle feature that is hard to capture from a theoretical perspective, it is of near negligible importance for any physical properties of the system. 
This is similar to the role of the local field correction (or, equivalently, the static XC-kernel) at high densities and temperatures~\cite{Dornheim_2020_PPCF}, which is not accurately captured by common dielectric theories; at the same time, physical observables such as the linear density response function are described with high accuracy even on the mean-field level in this regime.

\begin{figure*}
\center
\includegraphics[width=16cm]{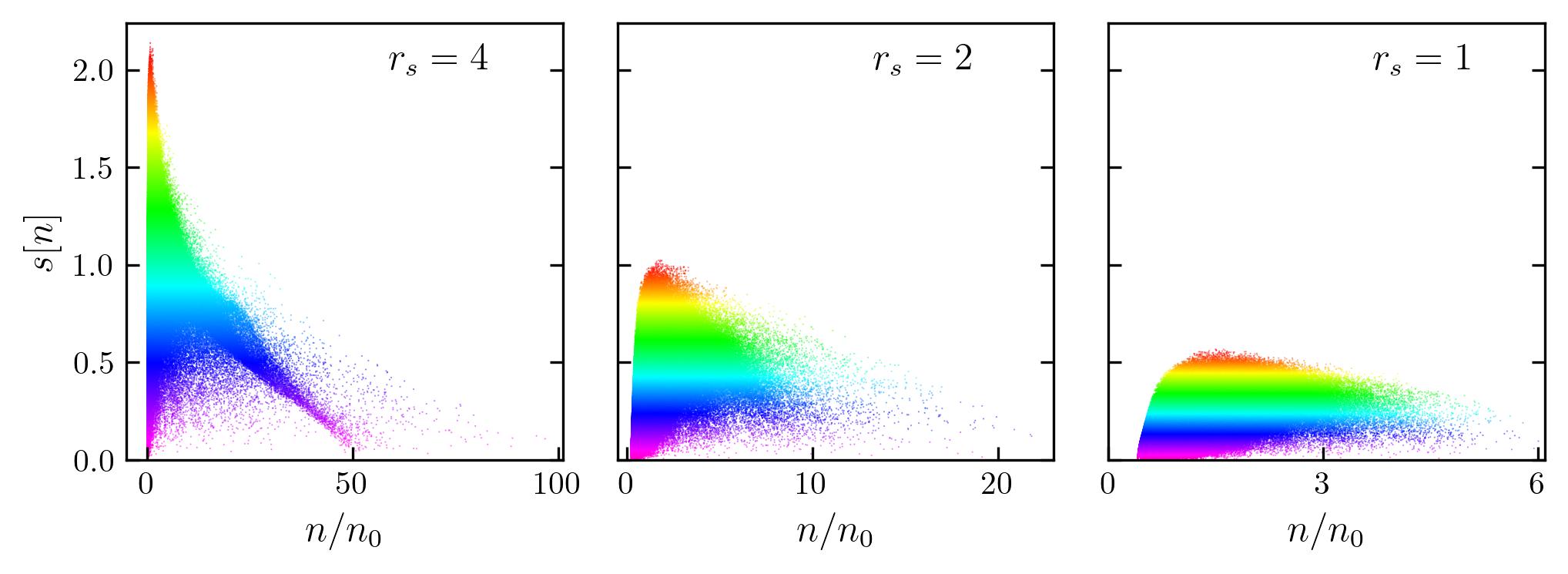}
\caption{ \label{fig:rdg_N500} 
Distribution of the RDG [Eq.~(\ref{eq:s})] with respect to density for warm dense hydrogen computed using KS-DFT data for 500 atoms using the T-LDA~\cite{groth_prl17} XC functional. Snapshots for $r_s=4$, $r_s=2$, and $r_s=1$ were generated by self-consistent KS-DFT-MD simulations at the corresponding densities and temperatures. 
}
\end{figure*}

\subsection{Utility of the RDG measure for DFT-MD simulations of bulk hydrogen}\label{s:DFT-MD}

To unambiguously demonstrate the general utility of the RDG for the detection of bound state breaking in warm dense hydrogen, we have carried out additional extensive simulations with $N=500$ hydrogen atoms. 
The snapshots are generated by performing KS-DFT-MD simulations for $r_s=1$, $r_s=2$, and $r_s=4$ at the electronic Fermi temperature, $T=T_\textnormal{F}$. 
In contrast to the previous example, we use snapshots that are self-consistently computed using  KS-DFT-MD independently for each set of parameters.
The results for the dependence of the RDG on the density are presented in Fig. \ref{fig:rdg_N500} for $r_s=4$ (left), $r_s=2$ (center) and $r_s=1$ (right). 
Clearly, we find the same qualitative trends as for the smaller synthetic snapshots analyzed in Fig.~\ref{fig:RDG_rs4} above.
Specifically, we observe a sharp peak of the RDG  at $r_s=4$ (with the maximum around ${\rm max} (s[n])\simeq 2.25$ and $\alpha_{\rm min}=0.0425$), which is a signal of the presence of bound states in the system. When the density parameter is decreased to $r_s=2$  and $r_s=1$, the sharply peaked structure disappears as the bound states are being broken. 
This is accompanied by  a significant decrease of the maximum value of the RDG distribution  with respect to the density.
At $r_s=2$ we have $s[n]\lesssim 1$ with $\alpha_{\rm min}=-0.025$ and at  $r_s=1$ we have $s[n]\lesssim 0.5$ with $\alpha_{\rm min}=-0.45$.
The presented results for 500 particles thus further demonstrate the utility of the RDG for the detection of bound state breaking in warm dense hydrogen.


\section{Discussion}\label{sec:outlook}

The formation and, conversely, breaking of bound states in hydrogen and heavier elements has an important role in HED physics; it is directly relevant both for experiments, e.g.~in the context of ICF, and for our understanding of astrophysical objects such as Jupiter.
Yet, the extreme conditions in HED experiments constitute a significant obstacle for the experimental study of this process. 
In this regard, \textit{ab initio} simulations are indispensable to understand the physics and chemistry  at these conditions. 
In practice, the most widely used first-principles method for such studies is KS-DFT. Having originally being developed for applications at ambient conditions,
the extension of KS-DFT to high temperatures remains significantly less developed compared to the ground-state case. 
In particular, the accuracy of a KS-DFT simulation decisively depends on the utilized XC functional.
Being of an a-priori unknown quality, new XC functionals~\cite{karasiev_gga_18,thermal_PBE_2023} require the rigorous benchmarking against test sets that are based on either experimental observations or higher level  simulation  methods.
In the present work, we present the first suitable test set for a real warm-dense matter system based on exact PIMC calculations.
It is freely available online~\cite{data} and can be used to benchmark XC-functionals, thereby potentially resolving the discrepancy between the thermal GGA functionals introduced by two independent groups in Ref.~\cite{thermal_PBE_2023} and Ref.~\cite{karasiev_gga_18}. Moreover, these data will be useful to guide the development of new tools for HED theory.



Among the presented findings, we first highlight the utility of the dimensionless RDG $s[n]$ in identifying the pressure induced ionisation in the medium.
Second, the generalized RDG $s[n]\left(n/n_0\right)^{\alpha}$ that has been introduced in this work can be used as a tool for the study of the interstitial electronic structure. 
Third, our analysis clearly highlights the 
importance of explicitly thermal contributions to the XC functional for the description of the density gradients.
It is most likely a direct consequence of the entropic contribution to the kinetic part of the total XC free energy that is completely missing from ground-state functionals.
In this context, we reiterate the relevance of information about the distribution of the RDG for the further development of thermal XC functionals, for example for ICF applications.

We are convinced that our work opens up a number of new avenues for impactful future research. First and foremost, our findings will inform the development of novel functionals that are specifically designed for applications in the HED regime. In this regard, a particularly promising route is given by a new, fully nonlocal class of thermal XC functionals based on a combination of the adiabatic connection formula with the fluctuation--dissipation theorem~\cite{pribram}. This line of research may eventually lift thermal KS-DFT calculations onto the same level of accuracy as KS-DFT calculations at ambient conditions, where its success regarding the description of real materials arguably remains unrivaled.

In addition to its demonstrated utility for the assessment of XC functionals for KS-DFT, the presented test set constitutes an unassailable benchmark for any theoretical method that is available in the warm-dense matter regime. This is particularly interesting to understand the accuracy of the fixed-node approximation in PIMC simulations. On the one hand, this restricted PIMC method~\cite{Ceperley1991} allows one to completely circumvent the exponential computational bottleneck due to the fermion sign problem~\cite{dornheim_sign_problem}, which has allowed Militzer \textit{et al.}~\cite{Militzer_PRL_2012,Militzer_PRE_2018} to present extensive numerical results for a gamut of WDM systems including second-row elements and composite materials~\cite{PhysRevE.103.013203}.
On the other hand, this paradigm constitutes a de-facto uncontrolled approximation in practice, and rigorous benchmarks have been hitherto limited to the comparably simple uniform electron gas model~\cite{Schoof_PRL_2015,Dornheim_PRB_nk_2021,Malone_PRL_2016}; here systematic errors of the order of $\sim10\%$ have been reported by Schoof \textit{et al.}~\cite{Schoof_PRL_2015}.
Our new test set thus presents the first opportunity to assess the accuracy of the fixed-node approximation in PIMC for a real warm-dense matter system.

We mention that the present study of the RDG can be extended beyond hydrogen, which has a particular relevance for HED science. More specifically, the complex interplay of numerous physical effects at these conditions leads to interesting effects such as partial ionization, in particular of heavier elements. Checking the capability of the RDG to capture the nontrivial superposition of different charge states thus constitutes a natural follow-up project of this work. 
Indeed, this line of thought might, for example, help to shine light onto the delocalization of atomic orbitals in Be at extreme temperature and density that has very recently been observed by D\"oppner \textit{et al.}~\cite{Tilo_Nature_2023} at the NIF.

Finally, we note that the RDG, generalised RDG, and the integrated measure Eq.~(\ref{eq:integrated}) can, in principle, be employed to analyse the electronic structure not only in materials at high temperatures, but also at low temperatures.


\appendix

\section*{Appendix}\label{sec:methods}

\subsection*{PIMC simulation details}

We have carried out \textit{ab initio} PIMC calculations to get exact results for the $3D$ single-electron density profile using the set-up described in Refs.~\cite{Dornheim_nanoscale_2023,Boehme_Folgepaper}. We use $P=500$ imaginary-time propagators within the pair approximation~\cite{MILITZER201688}, and the convergence with $P$ has been carefully checked. Note that we do not impose any nodal restrictions~\cite{Ceperley1991} on the thermal density matrix. Therefore, our calculations are computationally involved (we use $\mathcal{O}\left(10^5\right)\,$CPUh for a single calculation) due to the fermion sign problem~\cite{dornheim_sign_problem}, but exact within the given Monte Carlo error bars. The PIMC data are freely available online~\cite{data}, and can be used for a variety of applications.

\subsection*{Thermal KS-DFT simulation details}

The results from thermal KS-DFT simulations were computed using the GPAW~\cite{GPAW1,GPAW2} electronic structure code. 
For $r_s=4$, we used a standard PAW setup provided by GPAW to accelerate the convergence of our results~\cite{Lejaeghere2014Review}. 
For $r_s=4$, we set the $k$-points sampling to $10\times10\times10$, the energy cutoff to $440\,$eV and the main simulation cell size to $15.54~{\rm a_{0}}$ (with ${\rm a_{0}}$ being the first Bohr radius).
For $r_s=2$ and $r_s=1$, we have performed all-electron calculations with the field of each ion being represented by a bare Coulomb interaction. This is done to avoid potential problems related to the finite effective cutoff range at small distances in PAW setups. For $r_s=2$ and $r_s=1$, we used the energy cutoff $2700~{\rm eV}$ and k-points sampling $10\times10\times10$.
We set the number of bands for $N=14$  particles to be $N_b=280$ at all considered densities. 
The main simulation cell size at $r_s=2$ ($r_s=1$) is $7.77~{\rm a_{0}}$ ($3.885~{\rm a_{0}}$).
For occupation numbers smearing we have $T=3.132~{\rm eV}$ at $r_s=4$, $T=12.528~{\rm eV}$ at $r_s=2$, and $T=50.112~{\rm eV}$ at $r_s=1$.
As a convergence criterion for the self-consistent field cycle we required the change in energy  in the last three iterations to be less than $0.5~{\rm meV}$ per electron, the change in density to be less than $10^{-4}$  per electron,
and for the eigenstates the integrated value of the square of the residuals of the KS equations to be less than $4\times 10^{-8}~{\rm eV^2}$  per electron.
These simulation parameters have been rigorously tested with respect to convergence of the total energy.

For the RDG calculations of the system with $N=500$ particles at the $\Gamma$ point and using T-LDA \cite{groth_prl17}, we set the number of bands to $N_b=6000$ at $r_s=1$ ($T=50.112~{\rm eV}$) and $r_s=2$ ($T=12.528~{\rm eV}$), and to $N_b=2700$ at $r_s=4$ ($T=3.132~{\rm eV}$).
We performed KS-DFT-MD simulations using
the  GPU implementation of the Vienna ab initio simulation package (VASP) \cite{PhysRevB.54.11169, https://doi.org/10.1002/jcc.23096, HUTCHINSON20121422}.
To accelerate the equilibration part of the MD simulations, we used the hybrid approach reported in Ref. \cite{Fiedler_PRR_2022}. Accordingly, the KS-DFT-MD simulations where initialised using ionic coordinates from snapshots generated by performing orbital-free DFT-MD using the DFTpy package \cite{dftpy}.
Equilibrated snapshots were obtained from KS-DFT-MD simulations with 480 MD steps at $r_s=1$,  700 MD steps at $r_s=2$, and 3200 MD steps at $r_s=4$ (with each MD time step set to 0.01 fs).
The KS-DFT-MD simulations were performed using a Nose-Hoover thermostat with  the coupling  parameter of the system to the heat bath $Q$ (effective mass) set to $Q=0.5$.

\section*{Data Availability}
The data supporting the findings of this study are available on the Rossendorf Data Repository (RODARE)~\cite{data}.

\section*{Acknowledgments}
We gratefully acknowledge helpful comments by David Blaschke.
This work was partially supported by the Center for Advanced Systems Understanding (CASUS) which is financed by Germany’s Federal Ministry of Education and Research (BMBF) and by the Saxon state government out of the State budget approved by the Saxon State Parliament.
This work has received funding from the European Research Council (ERC) under the European Union’s Horizon 2022 research and innovation programme
(Grant agreement No. 101076233, "PREXTREME").
The PIMC calculations were carried out at the Norddeutscher Verbund f\"ur Hoch- und H\"ochstleistungsrechnen (HLRN) under grant shp00026 and on a Bull Cluster at the Center for Information Services and High Performance Computing (ZIH) at Technische Universit\"at Dresden.
This work was partially performed on the
HoreKa supercomputer funded by the Ministry of Science, Research and the Arts Baden-W\"urttemberg and
by the Federal Ministry of Education and Research, and
at the Norddeutscher Verbund f\"ur Hoch- und H\"ochstleistungsrechnen (HLRN) under grant mvp00024.
 MP and XS acknowledge the National Science Foundation under Grants No.\ CHE-2154760, OAC-1931473.
 

\bibliography{sn-bibliography}


\end{document}